\documentclass[letterpaper]{article} % DO NOT CHANGE THIS
\usepackage{aaai24}  % DO NOT CHANGE THIS
\usepackage{times}  % DO NOT CHANGE THIS
\usepackage{helvet}  % DO NOT CHANGE THIS
\usepackage{courier}  % DO NOT CHANGE THIS
\usepackage[hyphens]{url}  % DO NOT CHANGE THIS
\usepackage{graphicx} % DO NOT CHANGE THIS
\usepackage{natbib}  % DO NOT CHANGE THIS AND DO NOT ADD ANY OPTIONS TO IT
\usepackage{caption} % DO NOT CHANGE THIS AND DO NOT ADD ANY OPTIONS TO IT
\usepackage{algorithm}
\usepackage{algorithmic}
\usepackage{soul}
\usepackage[dvipsnames]{xcolor}

\usepackage{newfloat}
\usepackage{listings}
\DeclareCaptionStyle{ruled}{labelfont=normalfont,labelsep=colon,strut=off} % DO NOT CHANGE THIS
\floatstyle{ruled}
\newfloat{listing}{tb}{lst}{}
\floatname{listing}{Listing}
\nocopyright 
\begin{document}
%-------------------------------------------------------------------------------

%don't want date printed
\date{}

\title{\Large \textbf{You Still See Me: How Data Protection Supports the Architecture of AI Surveillance}}

\author{
Rui-Jie Yew\textsuperscript{\rm 1}, Lucy Qin\textsuperscript{\rm 2}, Suresh Venkatasubramanian\textsuperscript{\rm 1}
}
\affiliations {
  \textsuperscript{\rm 1}Center for Tech Responsibility,\\Brown University\\ \texttt{\{rui-jie\_yew,suresh\}@brown.edu}\\
  \textsuperscript{\rm 2}Georgetown University \\ \texttt{lucy.qin@georgetown.edu}
  }
\maketitle

\section{Abstract}
Data forms the backbone of artificial intelligence (AI). Privacy and data protection laws thus have strong bearing on AI systems. Shielded by the rhetoric of compliance with data protection and privacy regulations, privacy-preserving techniques have enabled the extraction of more and new forms of data. We illustrate how the application of privacy-preserving techniques in the development of AI systems--from private set intersection as part of dataset curation to homomorphic encryption and federated learning as part of model computation--can further support surveillance infrastructure under the guise of regulatory permissibility. Finally, we propose technology and policy strategies to evaluate privacy-preserving techniques in light of the protections they actually confer. We conclude by highlighting the role that technologists could play in devising policies that combat surveillance AI technologies.
%-------------------------------------------------------------------------------
\section{Introduction}

 Data forms the backbone of artificial intelligence (AI)~\cite{kalluri2023surveillance}. Data protection and privacy regulations thus have strong influence on AI development. Many current approaches to data protection and privacy laws impose requirements on the \emph{processing} (typically involving the collection  or possession\footnote{This is the case for Illinois BIPA~\cite{illinoisbipa}, which has seen some of the largest settlement amounts for computer vision AI technologies~\cite{yew2022regulating}, and other state biometric information privacy laws.}) of \emph{personal data}. Personal data -- data that is covered under these regulations -- typically includes data that is \emph{identifying}\footnote{The U.S. Video Privacy Protection Act defines personal information as ``information which identifies a person.'' Many data breach laws also employ this definition~\cite{schwartz2011pii}. The CCPA~\cite{ccpa}, for example, embeds a private right of action in the case of an information security breach and when the data was ``non-redacted'' or ``unencrypted''--following the identifying data model for the provision.}, \emph{identifiable}\footnote{Identifiable is a stronger definition than identifying, because it includes all the ways data can be combined to identify an individual. The EU GDPR adopts this definition for personal data, and \cite{greenleaf2020california} notes that this is a dominant approach globally in the definition of personal data.}, or \emph{sensitive}\footnote{Article 6 of the GDPR, for example, lays out a set of requirements for the lawful processing of \emph{personal} data, with additional requirements when \emph{sensitive} personal data  is processed. Japan and Colombia also have laws surrounding sensitive data, with their definitions emphasizing individual data that can be used to discriminate~\cite{solove2023data}.}. The greater the identifiability of data, the more onerous the requirements accompanying its collection and use~\cite{schwartz2011pii, zarsky2019privacy}. On the other hand, ``rights hinging on personal data being processed typically end when such data no longer becomes personal, falling out of scope of the relevant law''~\cite{veale2023rights}. Given the data-hungry nature of AI development, organizations are incentivized to keep their data out of legal scope. Even while the exact scope of covered data under data protection laws remains contested~\cite{finck2020they}, privacy-preserving techniques for AI systems continue to be researched~\cite{cummings2018role, treiber2022data} and wielded\footnote{``Federated learning is an innovative approach to machine learning for compliance. It enables multiple organizations to come together and train better quality models, while helping them to achieve their respective data privacy and security standards.''~\cite{microsoftcompliance2023}.} for legal compliance.

There is a significant body of scholarship in the AI community that discusses trade-offs and tensions between privacy, fairness~\cite{chang2021privacy, xiang2022being}, and accuracy/utility~\cite{suriyakumar2021chasing}. In these works, the analysis is typically focused on the effects of applying privacy-preserving techniques to fairness or system performance. \citet{xiang2022being}, in particular,  discusses a fundamental tension between being ``seen'' and being ``mis-seen'' in AI development for computer vision. Being ``seen'' is defined as ``having images of your face and/or body collected and processed for \textit{developing} HCCV [human-centric computer vision] systems'' and being ``mis-seen'' refers to ``experiencing poor performance from a
deployed HCCV system''. But, with the application of privacy-preserving techniques, data subjects may technically be ``\textit{un}-seen'' in the development of AI systems, only to have these systems turn around to take a good look.\footnote{As \citet{balsa2023pets} describe, a crucial aspect of the appeal of privacy-preserving techniques is their ``positive-sum quality: rather than imposing a trade-off between functionality and privacy, PETs [privacy-enhancing technologies] are hailed as a transformative way to protect privacy without compromising functionality.'' }

In this paper, we focus on how  privacy-preserving AI (PPAI) can be architected to take advantage of carve-outs in data protection and privacy laws and bolster surveillance infrastructure. A defining characteristic of surveillance is the power differential between those who see and those who are seen. Those in power are then enabled to ``see without being seen''~\cite{foucault2012discipline}. \textbf{Our goal is to illustrate how the cover of compliance with data protection and privacy laws can enable the use of PPAI to increase data consolidation and surveillance.} To this end, we examine the use of these techniques as part of three stages of system development and application: (1) dataset curation, (2) model computation, and (3) model application.

% Our goal is to illustrate how the rhetoric of compliance with data protection and privacy laws associated with PPAI blurs the eye of data protection and privacy laws even as PPAI can enable more and finer-grained consolidations human data.

First, we examine the use of secure multi-party computation (MPC) for dataset alignment~\cite{liu2023vertical} as part of dataset curation. MPC enables entities to combine information previously held in different silos across different contexts. We dig into Google and MasterCard's private data deal and note how MPC enabled the combination of user data from both entities, while being shielded by the rhetorical cover that the data involved was outside the scope of  privacy and data protection laws.

Second, we consider the uses of federated learning (FL) and homomorphic encryption (HE) as part of model computation. In the security literature, these techniques are avenues toward greater user control over their data and can protect against third-party intrusions or breaches. At the same time, they can also be applied to subvert relationship structures for data handling imposed by models of data protection, namely the controller-processor-subject model of data protection. We explain how the application of FL and HE can challenge conceptions of liability for data exchanges, resulting in scenarios where either \emph{no} entity or \emph{the users themselves} may be considered responsible for data collection. As a consequence, entities that use data resulting from FL and HE for model training may not be held accountable for model computations -- creating opportunities for more data accumulation. 

Third, we highlight the intrusions that system subjects might experience through the application of AI systems even when these subjects are not included as part of training data. When otherwise unavailable data is obtained through the use of privacy-preserving techniques, we question the grounds of the technical privacy-utility trade-off~\cite{ponomareva2023dp} for anonymization during model application and inference. The hallmark of a good privacy-preserving mechanism is one that straddles this trade-off--conferring the right amount of privacy protection while maintaining utility. However, the inherent relationality~\cite{viljoen2021relational, gordon2022case} of data could result in instances where high-level patterns in the population gleaned from data can in fact be better used to target individuals.

With data accumulation  at the core of how the AI pipeline is configured~\cite{kalluri2023surveillance}, there are incentives for privacy-preserving techniques to be used in AI development to get off the hook of data protection and privacy laws and to ultimately support infrastructures of surveillance. We propose technology and policy strategies to evaluate these techniques in light of the protections they actually confer. To conclude, we appeal to the role security researchers and technologists may play in developing regulations that implicate privacy-preserving techniques. The adversarial mindset crucial to assessing the utility of privacy-preserving techniques may also play an important role in modelling, mitigating, and shaping the effects of policies on technology development.

\section{Definitions and Background}

\subsection{Privacy-Preserving AI} 
Privacy-preserving artificial intelligence (PPAI) provides a set of techniques for AI systems that aim to protect against model \emph{attacks} and \emph{leakages} that extract information about the training data or model parameters~\cite{papernot2018sok, brown2022does}.
Privacy protections and their goals depend on the stage of AI development at which they are applied~\cite{xu2021privacy,ponomareva2023dp}. We provide a broad overview of different PPAI techniques with an emphasis on the different protections they provide as part of the layers of AI development. For a more comprehensive overview see, e.g., \citet{xu2021privacy}. We emphasize that the techniques we discuss have proven security guarantees that are the subject of extensive scholarship. Our discussion centers on the inadequacy of data protection frameworks rather than any inherent flaw to the underlying techniques.

\paragraph{Dataset Curation.}
As part of dataset curation, common privacy-preserving techniques include data sanitization~\cite{askari2012information} and differential privacy (DP)~\cite{brown2022does}. A sanitization mechanism aims to modify a given dataset to make it difficult for an adversary to confidently verify sensitive information about subjects in the dataset~\cite{askari2012information}. When applied at the dataset stage, DP can protect privacy with respect to a particular unit of information, such as the unit of words, phrases, or sentences~\cite{brown2022does}.

Datasets can also be constructed jointly by multiple entities through the use of MPC. MPC can be executed as part of private entity alignment~\cite{liu2023vertical}, which allows multiple parties to find intersections in their datasets and conduct joint analysis in a manner that does not reveal the original datasets to others. 

\paragraph{Model Training and Computation.}
As part of model training and computation, privacy protections may be implemented in various ways. For instance, as part of the process of model training, FL  enables multiple devices to jointly train a model without sharing that data to a central server~\cite{mammen2021federated}. It falls under an umbrella of techniques that processes and computes data locally such that the data never needs to be transfered to a company server. HE is another technique that can be applied at this stage. The application of HE enables model training to occur on encrypted data. This allows plaintext data to remain secure (not decrypted) even during model training.  

DP can also applied at this stage through the addition of noise to intermediate gradients~\cite{ponomareva2023dp}. This process controls for the influence that a piece of data has on the resulting model. Additionally, FL and DP can be simultaneously applied at this stage, where intermediate model updates are performed locally and differentially private noise is added during the aggregation of the updates. 

\paragraph{Using PPAI: Model Application.}
Straddling the trade-off between privacy and utility or accuracy is important for applying privacy-preserving techniques. This is commonly referred to as the ``privacy-utility trade-off''~\cite{zhong2022privacy}. Typically, the more privacy that is conferred through an application of a particular technique, the less utility the resulting model has. This is especially the case in data publishing, where additional noise directly compromises the accuracy of the published data.

There are also privacy implications that arise from the use of AI to perform inferences on existing data --- for instance, through the use of language models to infer additional information about a piece of text~\cite{staab2023beyond}. These implications are further discussed within the context of the political economy in the work of~\citet{solow2022information}.

\subsection{Policy Analysis: PPAI as Compliance with Data Protection and Privacy Laws}
Data protection and privacy policies mandate a staircase of requirements for data depending on its collection mechanisms and identifiability. Personally identifying or identifiable information opens the gate to the application of numerous privacy requirements in the United States.\footnote{For instance, Schwartz~\cite{schwartz2011pii} discusses how federal and state statutes rely on this distinction in its application --- including the Children's Online Privacy Protection Act, the Video Privacy Protection Act, as well as state security data breach notification statutes.} The EU's approach to data protection, most prominently the GDPR~\cite{eugdpr}, also relies on staggering protections to data: most requirements fall on the processing of personal data, then pseudonymous data, then anonymous data. This staggering of protections are also present as part of approaches to data protection in Japan and Colombia, which each have policies that pay additional attention to sensitive data~\footnote{See footnote 37 in \cite{solove2023data}.}. 

On the flip side, requirements lessen when data is no longer considered sensitive or personal --- typically, when data is considered pseudonymous or anonymous. For instance, the EU GDPR provides a safe harbor for anonymized data and lessens requirements for pseudonymous data. Across proposed state privacy and data protection legislation in the United States, data that is de-identified may also enjoy a safe harbor.
 
At the same time, requirements for pseudonymization and anonymization, or, whether a piece of data is no longer rendered personally identifying or identifiable, are not completely set in stone. For instance, \citet{burt2021guide} note that, while anonymization confers significant compliance advantages under the GDPR, a degree of regulatory uncertainty still remains surrounding the term itself. There  also remains uncertainty regarding how specific techniques might be used to anonymize data. The EU Article 29 Data Protection Working Party~\cite{wp} explains that there is no \emph{one} technique whose application confers data completely anonymized. Yet, regulatory and technical ambiguity surrounding definitions of covered data has not stopped companies from architecting PPAI to fall into the data protection and privacy's legal exceptions. As \citet{pollman2019tech} writes: ``The tech industry is notorious for design that pushes the regulatory envelope and aggressively uses rhetoric to defy legal norms and shape legal classifications.''

\subsection{Surveillance}
Surveillance has been described as ``collection and storage of information (presumed to be useful) about people or objects''~\cite{dandeker1990surveillance}, as well as the ``the
focused, systematic and routine attention to personal details for purposes of influence, management, protection or direction''~\cite{lyon2007surveillance}. \textbf{A key condition in surveillance relationships is the power differential between the surveilling and the surveilled --- that it ``must see without being seen'' or that one is ``totally seen, without ever seeing''.} ``He is seen, but he does see; he is the object of information, never a subject in communication''~\cite{foucault2012discipline, fuchs2010can}. 

%\lqnote{Want to add this point in here about the existing tie between the field of AI and surveillance, feel free to rephrase:}
We briefly note the historical context in which the field of AI (and in particular, computer vision) has been intrinsically tied to producing surveillance technologies.
An empirical analysis of academic publications in computer vision found that the vast majority of research in this area explicitly focused on extracting data about the physical body parts and movements of humans and that when patented, contribute overtly to systems for surveillance~\cite{kalluri2023surveillance}. The authors point to a pervasive trend across different types of institutions and an academic culture that has normalized such use cases, particularly through language that abstracts such use cases as to render them more neutral and to decontextualize from their real world applications. Such norms are important to consider and may persist when discussing PPAI.

 \emph{More} data and \emph{more} resource-intensive computations are required to extract value from PPAI infrastructure. Thus, it is typically the case that only high-resource entities have access to the value of these computations. This further entrenches the power dynamics between those who see and those who are seen. A person's data may not be leaked such that their identity could be stolen or such that they could be identified by a person. However, the effects of these computations and data uses may be more diffuse. Surveillance has been tied to influence, discrimination, and long-term ``social sorting''~\cite{lyon2003surveillance} --- the reinforcement of social differences between groups.  

 Importantly, however, by relying on the proven security guarantees of techniques such as MPC, DP, and HE, PPAI can also be used in contexts outside of surveillance and in the public interest. In recent years, studies have used MPC to understand workplace inequities through computing aggregate metrics (e.g., sums, averages). A study that began in 2013, which has been deployed multiple times, uses aggregate wage data contributed by different institutions to assess wage gaps (e.g., based on race, gender) in the city of Boston~\cite{lucy_qin_usability_2019, lapets_accessible_2018}. A recent study, involved data from individuals across 54 museums and analyzed various metrics related to workplace culture, such as issues around diversity and retention~\cite{gabriel_kaptchuk_jen_benoit-bryan_kinan_dak_albab_mia_locks_mayank_varia_good_2024,jen_benoit-bryan_workplace_2023}. In both cases, it was made clear ahead of time the data that would be used for analysis and the exact computations that would be executed. Those involved had to \textit{explicitly contribute} this data for the specific purpose of these studies.

\subsection{Privacy and Contextual Integrity}
Highly relevant to our work, CI provides a framework for ``evaluat[ing] the appropriateness and legitimacy of norm-breaching information flows by considering their ethical and social implications. CI thus requires that information flows be justified by more than technical means of confidentiality, anonymity, or consent management.''~\cite{mcguigan2023private,nissenbaum2004privacy}. While we do not provide an in-depth analysis on intrusions that arise from PPAI in this paper (our goal is to detail the consequences of opportunities for increased data consolidation and tracking created by data protection and privacy carve-outs), there is work that delves further into PPAI and privacy intrusions through the lens of CI. For instance, \citet{balsa2023pets} contributes CI analysis to three different technological deployments that add ``weight to the raising tide of voices that warn against a myopic conceptualization of privacy...too narrowly focused on the protection of personal or sensitive data''. Similarly, \citet{mcguigan2023private} draws on CI as privacy to examine how ``privacy-preserving'' ad-tech ``not only fail to achieve meaningful privacy but also leverage privacy rhetoric to advance commercial interests''.
\section{Seeing Through the Data-Protected Layers of the PPAI Pipeline}\label{sec:ppml}

In this section, we show how the layering of privacy-preserving techniques in AI development as part of the privacy-preserving artificial intelligence (``PPAI'') pipeline can enable new forms of data consolidation. At each point, the choice of how privacy-preserving techniques are used can refine mechanisms by which individuals' information is consolidated and synthesized. 

We consider three stages of AI development and use: (1) dataset curation, (2) model training and computation, and (3) model updating and application. We acknowledge that guarantees inherent to certain privacy-preserving techniques may be broken through how these techniques are used or operationalized under certain contexts~\cite{debenedetti2023privacy,boenisch2023curious}. For instance, assumptions that are made as part of privacy-preserving techniques do not always square with how they are used as part of AI development and deployment. This can lead to the system being used to extract insight from data subjects--whether during training or at deployment.\footnote{For example, \citet{boenisch2023curious} show that models trained with federated learning can still leak information about the individual subjects during  model training. \citet{brown2022does} also note that DP makes assumptions about the structure of training data that natural language training data does not always have. DP requires a clear boundary over where the privacy protections are applied. Where the boundaries of protectable information lie in the context of natural language text may not always be so clear. When DP is applied in these cases, this can then leak secrets of the data subjects as part of model output.} We do not focus on the consequences of when these techniques fail to live up to their technical promise. We operate under the assumption that they are in fact applied as technically intended--in the cases that anonymization guarantees hold and when entities are truly never in possession of clients' personal data. \textbf{Hearkening back to the definitions of surveillance we present, we illustrate how PPAI can nevertheless employ the rhetorical cover of data protection and privacy compliance to ``see without being seen''.}

\subsection{Producing PPAI: Dataset Curation\\
\normalsize{(You see even more of me.)}}\label{sec:PSI}
\begin{figure}
    \centering
    \includegraphics[scale=0.2]{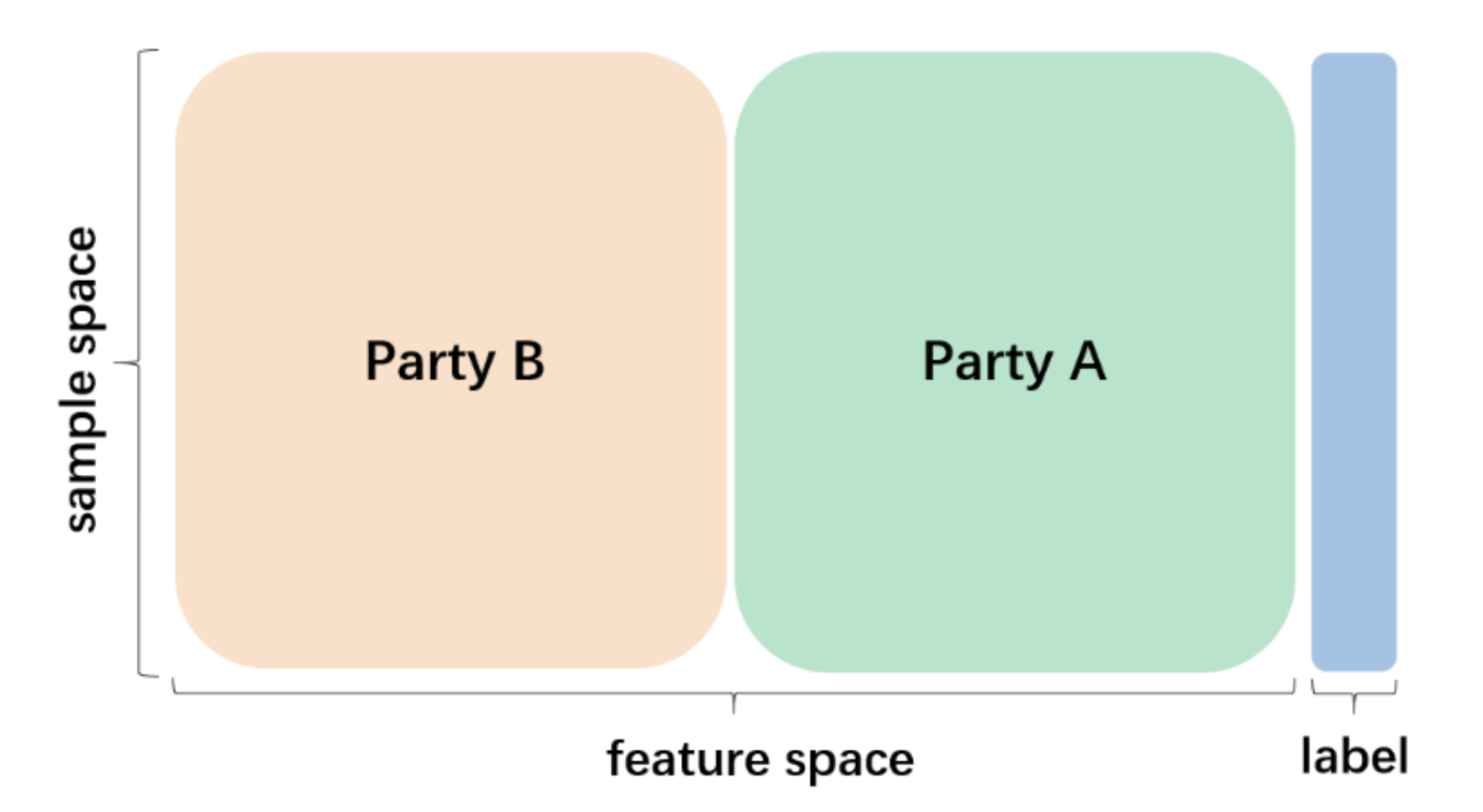}
    \caption{This illustration is from \cite{liu2023vertical}. It shows that Party A and Party B have the same samples, but different features of those samples. PSI, which is a pre-processing phase for vertical FL aligns the datasets from Party A and Party B to capture richer features for the samples that they share. Thus, for instance, if Party A has information that a data subject viewed an advertisement and Party B has information that a data subject bought the product served in the advertisement, then that information can be combined for training a resulting model.}
    \label{fig:psi}
\end{figure}

 In this section, we focus on how private, joint dataset curation can enable the creation of datasets with finer-grained information about each subject contained in them. We hone in on the use of secure MPC as part of this endeavor.\footnote{There are also a number of other techniques that can be applied to ML dataset creation for privacy preservation. When training language models, data sanitization is a common technique to remove identifying or sensitive information contained in the dataset~\cite{brown2022does}. For vision models, synthetic forms of data are increasingly developed and used for the training process~\cite{xin2022federated}. However, we focus on joint dataset creation, and, thus, MPC.} MPC is a cryptographic technique that can be used to compute a generic function using private inputs from different parties without revealing the inputs to any party~\cite{david_evans_pragmatic_2018}. This technique opens the pathway to computing on new forms of synthesized data.

 Private set intersection (PSI) is a specialized form of MPC that focuses on computing the intersection of two sets of private inputs. Given two entities that each hold an encrypted dataset, PSI can be used to compute the intersection of the two datasets without either entity ever needing to reveal their plaintext dataset to the other \cite{david_evans_pragmatic_2018}. This can then be used to combine datasets based on features, such as the intersection of individuals who live in the same zip code or have been to the same store between two datasets. This operation is illustrated in Figure~\ref{fig:psi}.

The use of MPC is heavily researched in the realm of finance and online advertising.\footnote{PSI and MPC have been proposed to train models using data from multiple sources to perform financial fraud detection~\cite{visasecure2022}. They have also been researched for the joint sharing of sensitive medical data to compute analytics and further medical research~\cite{kamm2013new}.} A central goal for advertisers and retailers is to accurately measure ad conversion (whether an ad led to an online or offline purchase). Security concerns make this linkage difficult because a user's plaintext data might be exposed across entities that may not have user permissions to access their plaintext data. Economic concerns also make the linkage of plaintext data difficult because human data is valuable, and companies do not have an incentive to share more data than what is profitable for them.

These concerns have sparked research in linking data across ad viewership and purchases so that involved entities do not have direct access to each other's plaintext data. In 2018, MasterCard sold Google customer data so that Google could understand how its ads affected in-person shopping.  A Google spokesperson described this method as a ``double-blind encryption technology that prevents both Google and our partners from viewing our respective users’ personally identifiable information'' ~\cite{mark_bergen_google_2018}.\footnote{In the same vein, as part of a talk at the Real World Cryptography conference in 2022, the head of Meta's private computation team cites increased data regulations as a reason for Meta's enthusiasm in building ``privacy-preserving'' ad conversion (which includes the use of PSI and DP) \cite{james_reyes_building_2022}.} The emphasis on a lack of access to identifying information can be important from a security standpoint. However, it, once again, potentially allows for these computations and data accumulations to fly under the radar of data protection regulations. For instance, \citet{walsh2022multi} note that ``most adoptions of MPC to date involve data that is not subject to privacy protection under the law.''\footnote{For a more nuanced analysis of how different data privacy laws might (not) apply to data used as part of MPC, see Section 4 in~\citet{walsh2022multi}.}  Even as the data involved may not be subject to legal protections, with a data-sharing method like MPC, not only do digital platforms who choose to use them have access to search histories and a user's ad views, they can then understand whether that data leads to \textit{what} someone buys in-person, \textit{when} they buy it, and \textit{how} much they spend. Its application provides further means of monetizing advertising and user spending data. The outcomes of these computations provide a new pathway for retailers and advertisers to move their tracking of consumer behavior beyond digital platforms into the physical world.

Resulting systems are referred to as privacy-preserving~\cite{kissner2005privacy} because the underlying dataset does not have to be revealed outright to the entities performing the joint computation. However, where economic and security concerns might have prevented this form of data synthesis in the past, PSI can be used to newly incorporate the link between a clicked advertisement and a payment, a doctor's visit and a person's dietary habits, whether done physically or online, enables a new and even more targeted form of individualization. The increased granularity of individual information through MPC in the resulting dataset can then set the stage for the rest of the AI pipeline as the data is ingested into further algorithmic training.

\subsection{Producing PPAI: Model Training and Computation\\
\normalsize{(You're protecting \emph{you}.)}}\label{sec:computation}
In this section, we consider how privacy-preserving model training and computation can be used to take advantage of relational structures imposed by data protection and privacy laws. By using certain privacy-preserving techniques, AI developers can position themselves outside the scope of accountability as they consolidate even more data.

There is  recognition within the privacy and security community that storing data on company servers leads to a loss of user control over how user data is processed and used--and, specifically, the data's vulnerability to external parties. As \citet{wang2019riverbed} note: ``unfortunately, there is a disadvantage to migrating application code and user data from a user’s local machine to a remote data center server: the user loses control over where her data is stored, how it is computed upon, and how the data (and its derivatives) are shared with other services.'' There have been two key responses to this problem from an AI privacy and security standpoint: (1) one approach has been to process user data locally or ``on-device'' and perform private immediate model training steps, typically through the use of techniques in on-device processing. This approach leverages FL~\cite{zhang2021survey} and intermediate steps may also be protected through the use of DP~\cite{stojkovic2022applied}, and (2) there has also been a significant amount of research in leveraging HE to compute models on encrypted data~\cite{heamazon2023}. 

Touching covered data through collection, processing, or possession triggers liability under data protection and privacy policies. These policies also impose a relational structure through which data is touched or exchanged.\footnote{The EU GDPR discusses controllers, processors, and subjects~\cite{eugdpr}. As another example, the CCPA adopts a relationship model involving businesses and service providers~\cite{ccpa}, but adopts similar definitions as the GDPR.} \citet{nissenbaum1996accountability} warned of ``the problem of many hands'' in complicating how accountability is determined from resulting problems. These methods, through exchanging data and computations in private forms through arguably no hands (in the case of FL) or many hands (in the case of HE), could deflect liability, at each point of data use for model training, for why intrusions that come from data processing are not theirs to handle.

Going back to the motivation for the adoption of these techniques in the security community, the potential deflection of responsibility can  make it difficult for users to exert control for how their data is processed as the data continues to accumulate during model training and computation.

As part of (1), FL is a popular privacy-preserving technique that allows computations for model training to be decentralized. Typically, FL coordinates training individually on user devices and model weights are then aggregated and sent to a centralized server to train a global model~\cite{konevcny2016federated}. Model updates can also be noised with DP before they are coordinated through a central server. However, when FL is treated as an avenue to more types of model computations, users may not have awareness over the fact that their data is being processed or how the resulting model is sent to corporate servers~\cite{veale2023rights}. From a legal standpoint, data subjects may also then unwittingly become data controllers or processors processing data on their own devices under the EU GDPR~\cite{veale2023rights}. Under a controller-processor-subject framework of data protection, controllers and processors owe responsibilities to data subjects. If data subjects are instead classified as controllers or processors, this could confound the attribution for harms resulting from and relating to data collection and processing~\cite{veale2023rights}--the users are collecting their own data!

As part of (2), HE has been offered up as a solution to model training that preserves the privacy of the training data as the model is being computed. Traditionally, computation on encrypted data requires data to be decrypted for the function to be applied. This leads to security vulnerabilities in the sense that users' plaintext data is revealed when the function is applied. In the case you are encrypting your own data to be computed, your plaintext data need not be revealed to any intermediary parties. \citet{acar2018survey} presents the example where you the data owner would like a third-party to compute a function on your message. The data owner would (a) encrypt the message, (b) send it to the server, (c) get the encrypted result of the computation back, and (d) decrypt the encrypted result. Where the data owner directly interfaces with the third-party, the data owner never reveals the message to anyone. 

In the ``confidential'' machine learning setting~\cite{sagar2021confidential}, however, model developers generally rely on the computational infrastructure of high-resource actors~\cite{dotan2020value}. There is a broader ecosystem of actors--some who may interface with plaintext data and some who may not. The model developer may have the dataset, but the model developer could then off-load model computation onto the cloud provider and the cloud provider only ever works with the encrypted dataset. Because only the encrypted dataset is computed on,  such a developmental process could result in a conceptualization where no one interfaced with data at all. 

Indeed, actors who provide computational infrastructure are already finger-pointing to take advantage of the EU GDPR's imposed relational structures. Several filed complaints in the EU detail how Microsoft labelled itself a ``processor'' of data and the school district a ``controller'' of data~\cite{noybmicrosoft2024}. Through this labelling, Microsoft is enabled to set the terms of computation and data storage while off-loading data auditing and transparency responsibilities onto the school districts--who often do not have the resources or expertise to conduct such tasks.
\subsection{Using PPAI: Model Application \\
\normalsize{(Don't use them against me!)}}\label{sec:updating}
In this section, we consider consequences of models that are produced from a privacy-preserving pipeline which are applied on subjects whose data may not necessarily be in the training dataset. Our discussion builds on and echoes the work of privacy scholars, who have long argued that collective insights nonetheless have leverage in targeting individuals~\cite{solow2022information, viljoen2021relational, solove2023data,barocas2014big}. 

\citet{dwork2011firm} highlights: ``the things that statistical databases are designed to teach can, sometimes indirectly, cause damage to an individual, even if this individual is not in the database.'' The influence of one person's data on another is increasingly recognized as a point that data protection laws have overlooked. The collective influence~\cite{ligett2023we} of data is made glaringly clear in the context of genetic data. In 2023, 23andMe announced that it had suffered a data breach. As part of the data breach's consequences of the ``DNA Relatives'' feature, this data breach had an effect not only the subjects directly affected by the breach, but also on their relatives who might not even be aware that 23andMe possesses their information~\cite{23andme2023}.

The collective influence of data is also reflected in how  technical operations are performed on it. In the case of recommender systems, the popular technique of collaborative filtering~\cite{su2009survey} explicitly leverages data about other people's online experiences to analyze how it can be used to shape \emph{yours}--weighting the preferences and experiences of those that are most similar to you the highest. Inferences are drawn from others to be made about you. Recent work has also pointed to the importance of shifting from understanding privacy in AI systems through the lens of identifiability to understanding privacy through the impacts of their inference. For instance, it has been shown that language models can be used to \emph{infer} sensitive information from text that may not appear to have any identifying information~\cite{staab2023beyond}. \citet{staab2023beyond} show that from a seemingly innocuous text passage: ``there is this nasty intersection on my commute, I always get stuck there waiting for a hook turn'', language models could infer that the author of the passage is likely from Melbourne because of their use of the term ``hook turn''.

This relationality~\cite{viljoen2021relational} inherent in both data itself and the techniques that use them to shape our online experiences challenge an individualistic conception of anonymity. 

A common trade-off when applying anonymization techniques is the privacy-utility  or the privacy-accuracy trade-off: how much utility or accuracy in the resulting data do you want to sacrifice to preserve the privacy of data subjects used to train the model?\footnote{ See, e.g.,~\cite{ponomareva2023dp}: ``Generally, there is a trade-off between $\epsilon$ and the utility of the mechanism (e.g., accuracy of a neural network); smaller $\epsilon$'s typically lead to lower utility if other variables like the dataset size and batch size remain constant.''} Typically, the more privacy protection a technique confers to data subjects, the less useful the result becomes--whether the result is a private statistic or a machine learning model. When privacy-preserving techniques are applied in practice, decisions must be set about the parameters that correspond to the degree of privacy protection. For example, in the application of differentially private algorithms, the value of $\epsilon$ must be carefully chosen so that results still have utility. The hallmark of a good privacy-preserving mechanism is one that straddles this trade-off--conferring privacy protection while maintaining good utility. 

However, there are instances where this trade-off may not even exist. The application of privacy-preserving techniques can \emph{help} learning~\cite{dwork2015generalization, nisthowto2021}. Privacy-preserving generalization from collective data can, in some instances, be even  better for learning. In other words, knowing about others \emph{like} you can help AI systems hone in on you. This begs the question of how privacy and utility should be formulated when anonymization techniques are applied. Particularly, there may be certain data uses toward the production of an AI system in which a large degree of utility can itself represent a intrusion no matter how individual privacy protection is conferred in the development of PPAI.

\section{Regulating PPAI: Technology and Policy Strategies}

PPAI can be used to extract value from previously untouched crevices of data while limiting liability for the consequences of those extractions. The layering of techniques at each stage of the development pipeline can lead to data infrastructures that bolster AI surveillance. In this section, we provide technical and policy strategies to address the regulation of these developments and uses. Crucially, we also consider how privacy-preserving techniques can themselves also be wielded to protect against surveillance. Requirements hinging on personal data continue to be at the core of upcoming privacy and data regulations with relevance to AI\footnote{The United States, for instance, will see data protection policies in Delaware~\cite{delawarepdpa}, New Jersey~\cite{newjerseysb332}, and Texas~\cite{texasdataprivacyandsecurityact} (among others) come into effect in the next few years.}, making these strategies urgent to consider.

\subsection{Regulating PPAI as Value Extraction}

The application of privacy-preserving techniques may be motivated by the \emph{value}~\cite{parsons2023valuing} of the resulting data that is gained through its application. Data value and the value of information that is not covered under approaches to data governance have been discussed in~\cite{viljoen2021relational,solow2022information}.

For example, \citet{jones2020nonrivalry} describe encryption as a technique that can be leveraged to control and monitor access to data as a resource. By leveraging techniques such as encryption, the economic value of data can be increased by selectively excluding access to data from others. Similarly, the use of DP in industry to access new kinds of datasets as discussed in~\citet{kearns2019ethical} increases value by allowing access to new kinds of data that could not be accessed before. The use of MPC can allow multiple firms to unlock the value of their data while simultaneously controlling for each other's direct data access.

\citet{parsons2023valuing} and \citet{viljoen2021relational} have discussed the value of data that is overlooked in the regulation of data privacy. In this section, we consider technical strategies that could leverage the principles of purpose limitation to control value extraction. Additionally, we consider policy strategies for capturing the power relationship firms have in the accumulation of collective data that impacts individuals.  

\paragraph{Purpose Limitation and Specification.} 

Purpose limitation is an oft-discussed privacy principle~\cite{forgo2017principle}. Yet, in the context of data protection, it is largely understood to only apply to personal data. For instance, purpose limitation has been described as a principle that personal data ``must be collected for specified, explicit and legitimate purposes and not further processed in a way incompatible with those purposes''~\cite{forgo2017principle}. 
However, value accumulates through the relational and social nature of data~\cite{parsons2023valuing}. System users also conceptualize privacy harms not in terms of identifiability but in terms of \emph{use}. The study by \citet{kacsmar_comprehension_2023} on user expectations on the use of methods like MPC, PSI, and DP found that there were concerns that the use of the data, regardless of whether in plaintext or encrypted form, could be inherently privacy-invasive. Notably, one of their study participants voiced that,\emph{ ``If you're using my data, then there's no privacy... if there's privacy, then you're not using my data."} 

The value of relational data may still be expressed in ways that are amenable to technical, cryptographic notions of purpose limitation--for instance, through limiting the amount of times that a particular piece of data is accessed, limiting the time period that data is made accessible, as well as through limiting the types computations that be made. 

There is a body of research on the development and use of ephemeral data structures or self-destructing data structures~\cite{perlman2005ephemerizer, roxana_geambasu_vanish_2009, zeng2010safevanish, seny_kamara_crypto_2020}, which could be helpful in the enforcement of purpose limitations along the axis of time. These proposals seek to guarantee time-based data deletion even when the data is stored by an untrusted entity that does not explicitly perform any deletion operations (e.g. deleting entries from a database). This is one direction of work that seeks to apply direct accountability and limitation to how data is accessed (and then used). Future research could explore similar solutions for PPAI that enshrine technical guarantees around the time period (or number of times) data can be used, or the data flows that surround the use of data for model training.

Another way purpose limitation could be implemented is through limiting the types of computations that can be performed on data. A few MPC-based systems for collaborative analytics on datasets from different sources have proposed integrating mechanisms for checking query compliance against a set of pre-defined policies \cite{de_viti_covault_2022, volgushev_conclave_2019, ben_getchell_ccd-mpc_2019}. For example, this could be used to limit \textit{any} computation on specified demographics to prevent targeting. This could also allow individual entities to set their own policies around how their data is used and computed on. Since this would require a preemptive specification of allowable computation, it may increase transparency around how data is used to help users make informed decisions. These ideas could similarly be adapted to PPAI systems, and there may also be privacy implications for the growing body of research on limiting dual uses of AI systems~\cite{henderson2023self}, particularly pre-trained foundation models.

\paragraph{A Duty to System Subjects: how \emph{our} data implicates \emph{my} system use.}

Data accrues value \emph{at scale} and \emph{in relation} to others~\cite{malmgren2019resisting}. Crucially, data accrued at scale and in relation to each other can also serve to refine the mechanisms through which to \emph{target specific individuals}. Yet, the communication of the mechanics and privacy guarantees of privacy-preserving techniques have focused on promises to individuals.\footnote{For example, descriptions of DP guarantees focus on identifying an individual in a database or dataset. But system subjects, who may or may not be in the dataset, are also implicated when anonymization techniques are used. } 

Because of the power that is inherent in architecting the relationship between data more broadly and user experiences, fiduciary privacy duties to personal data are an increasingly popular approach to regulate the accumulation of data in ways that intrude on privacy. However, linking these duties to an individual's personal data can neglect to account for how collective data is leveraged to shape individual's online experience. We propose that carve-outs from duties surrounding how personal data is processed and used should be additionally conditioned upon a duty to \emph{system subjects}.

Recent legal scholarship propose fiduciary duties~\cite{balkin2015information}--specifically, duties of loyalty~\cite{richards2021duty} and care~\cite{citron2022fight}--to address the privacy harms that persist in light of current models of data protection. These duties encompass a recognition of relationships that hold power imbalances, and its core principles can be extended across many different types of relationships, such as both interpersonal and professional relationships~\cite{miller2018identification}. A duty of loyalty's main principles include a prohibition of self-dealing~\cite{richards2021duty}, preventing conflicts of interests~\cite{gold2019fiduciary}, and duties to disclose information~\cite{gold2019fiduciary}. By capturing asymmetric power relations between data collectors and data subjects~\cite{richards2021duty}, these proposed duties represent an important step in preventing surveillant privacy harms to data subjects.\footnote{However, there remains contention about where a duty of loyalty fits in relationship to other duties and obligations~\cite{gold2019fiduciary, khan2019skeptical}. \cite{richards2021duty} argues that the possibility for ``conflicting loyalties'' are already embedded as part of a duty of loyalty and may be overcome through carefully scoping the duty.} However, proposed duties as part of privacy law are often still tethered to a data subject's \emph{personal} or \emph{identifying} information. Such is the case even in \citet{richards2021duty}'s proposals, which discusses a ``duty of loyalty for personal information''. In the proposed United States Data Care Act, each duty is scoped to individual identifying data or inferences that are drawn from individual identifying data~\cite{datacareact}.\footnote{The Data Care Act contains a duty of care, a duty of loyalty, and a duty of confidentiality. Each of these duties are scoped to ``individual identifying data'' or, in the case of a duty of loyalty, ``data derived from individual identifying data''.} A duty of loyalty is included as part of the ADPPA~\cite{adppa} emphasizes the minimization of covered data:  data that ``identifies or is linked or
reasonably linkable to an individual''~\cite{adppa}. This scoping can fail to capture surveillant intrusions when associated data that is not identifying or personal, such as through targeted profiling~\cite{koops2008some}. A duty of loyalty or care scoped this way implicitly links a particular individual's online experience with only \emph{their} respective collected data. But, as we discuss in Section~\ref{sec:updating}, data and the techniques that are used to extract value from data are inherently relational.

To capture the power relations that firms have in leveraging data (be they exclusively tied to an individual or collectively ours) to curate each individual's online experiences, it may be important to condition carve-outs for anonymized data upon a duty to system subjects (whether or not those system subjects are contained in the dataset) --- recognizing the power that others' data holds over individual experiences with online systems. This would include a recognition of not just how individual data is leveraged, but also how others' data is leveraged to impact individuals. To that end, we echo proposals that encourage a collective perspective into the effects of data.  \citet{gordon2022case} emphasize the importance of a collective perspective in understanding the ways that collective data is leveraged in the process of personalization. This  perspective is built through not just best practices or requirements in how data is collected, but through insight and monitoring of how that data is used~\cite{ligett2023we}. \citet{koops2008some} makes an argument in a similar spirit that the focus should be on how data profiles are \emph{used} rather than on the data pre-processing stages. But, perhaps contradictorily, this would require a level of information about user behavior and how this behavior triggers the content that is fed back to them. The use of privacy-preserving techniques can be incentivized toward this end. \citet{gordon2022case} highlight the role that DP can play in protecting the privacy of individuals in obtaining this insight. Research that supports the monitoring content presentation on  platforms~\cite{chen2023subscriptions} could also provide insight into both how user interactions prompt certain pieces of content as well as the types of content that are presented to many people as a whole. In the case of surveillant privacy harms built up by privacy-preserving techniques, data cooperatives could negotiate for increased transparency regarding how a collective body of data is concentrated to impact individuals.

Regulatory agencies and could also play an important role in mapping the trajectory of collective data. \citet{van2019regulatory} defines regulatory monitors as ``those whose core power is to regularly obtain nonpublic information from businesses outside the legal investigatory process.'' \citet{van2019regulatory} notes that, in addition to rule-making and enforcement, agencies increasingly play a role in monitoring and information collection.  A tighter feedback loop of insight into data use can provide incentives that preempt uses of privacy-preserving techniques to support surveillance infrastructure.

\paragraph{Communicating Privacy Expectations.} 
As part of a duty to system subjects, there is a need for accurate and clear communication on the privacy guarantees and purposes of PPAI. Many users have misconceptions about the core security guarantees of end-to-end encryption, which is commonly found in many messaging apps~\cite{abu-salma_obstacles_2017, abu-salma_exploring_2018, akgul_evaluating_2021, justin_hendrix_what_2023}. 
For tools that may be part of the PPAI ecosystem, studies have been conducted on explanations of DP and MPC along with user expectations of their privacy~\cite{ cummings_i_2021, agrawal_exploring_2021, corman_public_2022, karegar_exploring_2022, 
xiong_exploring_2023, nanayakkara_what_2023, kacsmar_comprehension_2023,}. 
Synthesizing these findings, communicating the technical guarantees of privacy-preserving tools and effectively creating accurate expectations of privacy to non-expert audiences is a nuanced process that can be difficult to get right\footnote{We refer readers to the literature cited on recommendations for communicating privacy guarantees.}. 
In addition, transparency is needed on the data used and the broader purpose of proposed systems. As \citet{kacsmar_comprehension_2023} found in their study on user perceptions of tools for secure computation (e.g., PSI, DP, MPC), these factors were just as critical as technical explanations when deeming whether they were appropriate for use.

Statements about the privacy guarantees of PPAI must therefore be carefully articulated so as to not overstate their protections. Rhetoric associating privacy-preserving tools with public interest and non-profit entities may create higher levels of trust and an increased willingness toward permitting data usage, regardless of the actual privacy protection \cite{wang_role_2024}. Accurate explanations and transparency about data usage are necessary as to not create false understandings of privacy protections as means of enabling new forms of data collection.

\subsection{Regulating PPAI as Liability Management}

PPAI may allow firms to construct barriers from condemning information and, thus, build boundaries against liability from that information. If platforms can claim that they do not possess fine-grained enough data to provide that information, or that the majority of that information is spun through local devices, they may be considered off the hook for downstream harms. As an example, to protect against being found to make biased or discriminatory decisions \cite{kumar2022equalizing, van2022privacy},  financial institutions have limited their collection of certain types of data under the guise of increased privacy. This has led to challenges to developing AI algorithms that account for impacts to individuals. Additionally, we describe how local processing or HE for model training can serve to  complicate liability attribution for resulting model characteristics and use. Indeed, privacy-preservation can be wielded as a pretext~\cite{van2022privacy} and as a form of liability shielding  rather than confer meaningful protections to associated individuals~\cite{pozen2016privacy}. This raises the question of how to promote incentives to use privacy-preserving techniques in ways that are protective to individuals while holding developers accountable for how data is used downstream. In this section, we consider standards of proof and the development of infrastructures for collective transparency.

\paragraph{Standards of proof.}

 In light of ways that privacy can be used as a pretext, \citet{van2022privacy} proposes that there should be higher standards of proof when data is considered withheld from regulators--requiring firms to motivate that the risk of privacy intrusion is great and truly requires that the data be withheld. Where privacy-preserving techniques are used, perhaps the more relevant question may be whether their uses truly create a barrier to the the investigation of data and system effects-- such as how the resulting system can be used to manipulate or discriminate against system subjects. There is a growing amount of research that addresses how auditing can be performed even through data protection~\cite{imana2023having}. 
 
 The way that privacy-preserving systems may be architected to shield against potentially condemning information hearken to dynamics in patent law, under which firms may try to design \emph{around} existing patents to prevent liability for infringement~\cite{burk2016perverse}. The doctrine of equivalents is the principle that ``a substitution that performs the same function
in the same way with the same result as a patented item, even if not
formally within the text of the claims, still infringes by equivalents,
if not literally''~\cite{burk2016perverse}. \citet{burk2016perverse} notes that the doctrine of equivalents helps to separate meaningful inventing around from ``wasteful'' inventing around. 

Similarly, standards for the use of privacy-preserving techniques could be developed to differentiate their use to protect individual privacy from their use to prevent liability for harmful impacts to individual users. An analogous standard might ask whether a particular ``privacy-preserving'' architecture confers meaningful individual protections in light of the ultimate purpose/use of the system.

\section{Adversarial Compliance}
Compliantly winding around rules while subverting their goals has been recognized in several ways. \emph{Avoision} ``describes conduct which seeks to exploit the differences between a law's goals and its self-defined limits''~\cite{giblin2014aereo, katz1996ill}. \citet{burk2016perverse} terms \emph{perverse innovation} as a form of technological development designed to avoid regulation in response to the inescapable existence of loopholes in regulation. \citet{schneier2023hacker} additionally describes this conduct as a sort of legal \emph{hack}-- specifically noting that, while hacks are traditionally connoted as a way for underdogs to fight against the powerful, hacks to the legal system are often exploited by those who are most powerful. Crucially: perverse innovation, avoision, and hacks are different from simply breaking a rule and thereby evading enforcement. They are ways that rules may be avoided altogether, hence bypassing any trigger of liability. Software and technology more broadly has historically played a key role in such conduct.\footnote{\citet{wu2003code} describes code and specifically the development of peer-to-peer networks as a way of bypassing and challenging copyright law altogether.} \citet{wu2003code} challenges the notion that code is law~\cite{lessig2000code} and proposes an alternative framing of code as a form of avoidance of the law.   This calls to mind the framing of anonymity presented in~\citet{barocas2014big}, that ``anonymity obliterates the link between data and a specific person not so much to protect privacy but, in a sense, to bypass it entirely.'' With data protection regulating data by tiers of identifiability and sensitivity, anonymization is a way that rules surrounding data protection may be bypassed.
 
 The properties of how this mismatch between anonymization techniques and data protection laws occur may, in part, be modelled by tools that are already familiar to security technologists. Security is often conceptualized as a never-ending cat-and-mouse game between systems builders and hackers. The legal system has been conceptualized as a similarly never-ending cat-and-mouse game between the regulator and the regulated. When technology plays a large role in how regulation is determined, so, too, may be the relationship between technologists and regulators. \citet{choi2005grokster}, for instance, discusses the relationship between copyright liability and the design of peer-to-peer networks as a cat-and-mouse game. With  each copyright verdict hinging on the design of peer-to-peer networks, their design quickly shifted to limit resulting liability--eventually leading to the ``Grokster dead-end''.

Adversarial thinking may be helpful to bring to bear the relationship between how privacy-preserving techniques extract value of data, a policy's intent, and how the written rules may or may not bring us to those goals. \citet{schneier2023hacker} similarly advocates for  red-teaming to test the robustness of laws in order to combat legal hacks. For instance, in the field of tax law, \citet{schneier2023hacker} suggests that tax lawyers be at the table when tax laws are designed with a ready eye for potential ``hacks'' or loopholes. Similarly, there may also be a place for privacy and security technologists in considering the design of data-protected systems in light of policy intentions. How might that technology be used to subvert the spirit of a particular regulation? Many research papers that architect privacy-preserving systems are explicitly motivated by compliance with data protection laws such as the GDPR~\cite{abbas2023exploring, jiang2021gdpr}. However, for the design of robust policies and systems that do respect user privacy, it may also be crucial to bring the adversarial mindset inherent to the security profession to the policy-making drawing board. 

\section{Conclusion}
Data protection and data privacy laws constitute an important prong of regulation because of their breadth. As long as personal data is processed or used, this body of law kicks in. However, on the other side of this breadth, is a potentially equally broad carve-out for when personal data is \emph{not} collected, processed, or used. If privacy-preserving techniques can run a truck through this carve-out, the development of PPAI can be developed to optimize for the use of data in ways that support individual intrusions and surveillance. Importantly, data protection law is not alone in how it is being challenged by the ways AI leverages and optimizes with collective data.  AI development challenges individualistic notions of data regulation across multiple levers.\footnote{In copyright law, for instance, the burden of proof lies on individual artists to defend their rights. Proposed legal evaluations of what it means for AI systems to copy a work during training hinge on a conception of the training process that takes specifically from individual works--when, in reality, generated works are impacted by multiple works, some that may even be unexpected to the human eye.} The development and application of data protection laws represent an important opportunity to conceptualize the role of data for AI systems more broadly.

 In this paper, we examine the interaction between the incentives put forth by data protection regulations and PPAI as a response. We show that the same techniques  incentivized to implementation by data protection and privacy laws--such as  MPC, FL, and HE--can be leveraged refine the AI surveillance pipeline.  Carve-outs for the use of privacy-preserving techniques hinge on an implicit promise to protect individuals. From the developments and uses we highlight, however, we draw out two overarching principles that PPAI  may be used toward instead--value extraction and liability management. We provide technology and policy strategies to think through how to regulate how PPAI are used as part of each. Finally, we highlight the role of technologists in thinking through the role of privacy-preserving techniques in regulatory compliance. 

\section{Positionality Statement}
We use ``artificial intelligence (AI)'' rather than ``machine learning (ML)'' to situate our work within the ongoing policy discussions surrounding the technology. This reflects our desire to convey the broad scope of the concerns we express. These concerns pertain not just to predictive technologies most closely associated with the term ML, but also the more generative AI frameworks with their insatiable desire for data.

Collectively, we are scholars of computer science, machine learning, technology policy, and cryptography. The first author has developed code for open-source DP libraries within an anonymization team. The second author has designed and deployed cryptographic systems that include the use of MPC. The third author has broad expertise in the deployment of AI systems in society, and the design of effective technology policy.

\section*{Acknowledgements}
We thank Kris Shrishak, Rebecca Spiewak, Alla Goldman Seiffert, Taylor Lynn Curtis, Gabe Kaptchuk, Zheng Dai, Sunoo Park, Victor Youdom Kemmoe, Palak Jain, and attendees at the 2024 Privacy Law Scholars Conference (PLSC) for helpful pointers and feedback.

Lucy Qin is funded through the Fritz Fellowship Program in Tech \& Society at Georgetown University. Rui-Jie Yew is funded through the Dihua Presidential Fellowship at Brown University.

\bibliography{refs/bibliography, refs/lucy}

\begin{thebibliography}{124}
\providecommand{\natexlab}[1]{#1}

\bibitem[{Abbas et~al.(2023)Abbas, Ahmad, Syed, Anjum, and Rehman}]{abbas2023exploring}
Abbas, Z.; Ahmad, S.~F.; Syed, M.~H.; Anjum, A.; and Rehman, S. 2023.
\newblock Exploring Deep Federated Learning for the Internet of Things: A GDPR-Compliant Architecture.
\newblock \emph{IEEE Access}.

\bibitem[{Abu-Salma et~al.(2018)Abu-Salma, Redmiles, Ur, and Wei}]{abu-salma_exploring_2018}
Abu-Salma, R.; Redmiles, E.~M.; Ur, B.; and Wei, M. 2018.
\newblock Exploring {User} {Mental} {Models} of {End}-to-{End} {Encrypted} {Communication} {Tools}.
\newblock In \emph{8th {USENIX} {Workshop} on {Free} and {Open} {Communications} on the {Internet} ({FOCI} 18)}. Baltimore, MD: USENIX Association.

\bibitem[{Abu-Salma et~al.(2017)Abu-Salma, Sasse, Bonneau, Danilova, Naiakshina, and Smith}]{abu-salma_obstacles_2017}
Abu-Salma, R.; Sasse, M.~A.; Bonneau, J.; Danilova, A.; Naiakshina, A.; and Smith, M. 2017.
\newblock Obstacles to the {Adoption} of {Secure} {Communication} {Tools}.
\newblock In \emph{2017 {IEEE} {Symposium} on {Security} and {Privacy} ({SP})}, 137--153. San Jose, CA, USA: IEEE.
\newblock ISBN 978-1-5090-5533-3.

\bibitem[{Acar et~al.(2018)Acar, Aksu, Uluagac, and Conti}]{acar2018survey}
Acar, A.; Aksu, H.; Uluagac, A.~S.; and Conti, M. 2018.
\newblock A survey on homomorphic encryption schemes: Theory and implementation.
\newblock \emph{ACM Computing Surveys (Csur)}, 51(4): 1--35.

\bibitem[{Agrawal et~al.(2021)Agrawal, Binns, Van~Kleek, Laine, and Shadbolt}]{agrawal_exploring_2021}
Agrawal, N.; Binns, R.; Van~Kleek, M.; Laine, K.; and Shadbolt, N. 2021.
\newblock Exploring {Design} and {Governance} {Challenges} in the {Development} of {Privacy}-{Preserving} {Computation}.
\newblock In \emph{Proceedings of the 2021 {CHI} {Conference} on {Human} {Factors} in {Computing} {Systems}}, 1--13. Yokohama Japan: ACM.
\newblock ISBN 978-1-4503-8096-6.

\bibitem[{Akgul et~al.(2021)Akgul, Bai, Das, and Mazurek}]{akgul_evaluating_2021}
Akgul, O.; Bai, W.; Das, S.; and Mazurek, M.~L. 2021.
\newblock Evaluating {In}-{Workflow} {Messages} for {Improving} {Mental} {Models} of {End}-to-{End} {Encryption}.
\newblock In \emph{30th {USENIX} {Security} {Symposium} ({USENIX} {Security} 21)}, 447--464. USENIX Association.
\newblock ISBN 978-1-939133-24-3.

\bibitem[{Amazon(2023)}]{heamazon2023}
Amazon. 2023.
\newblock Enable fully homomorphic encryption with Amazon SageMaker endpoints for secure, real-time inferencing.
\newblock \url{https://aws.amazon.com/blogs/machine-learning/enable-fully-homomorphic-encryption-with-amazon-sagemaker-endpoints-for-secure-real-time-inferencing/}.

\bibitem[{Askari, Safavi-Naini, and Barker(2012)}]{askari2012information}
Askari, M.; Safavi-Naini, R.; and Barker, K. 2012.
\newblock An information theoretic privacy and utility measure for data sanitization mechanisms.
\newblock In \emph{Proceedings of the second ACM conference on Data and Application Security and Privacy}, 283--294.

\bibitem[{Balkin(2015)}]{balkin2015information}
Balkin, J.~M. 2015.
\newblock Information fiduciaries and the first amendment.
\newblock \emph{UCDL Rev.}, 49: 1183.

\bibitem[{Balsa and Shvartzshnaider(2023)}]{balsa2023pets}
Balsa, E.; and Shvartzshnaider, Y. 2023.
\newblock When PETs misbehave: A Contextual Integrity analysis.
\newblock \emph{arXiv preprint arXiv:2312.02509}.

\bibitem[{Barocas and Nissenbaum(2014)}]{barocas2014big}
Barocas, S.; and Nissenbaum, H. 2014.
\newblock Big data’s end run around anonymity and consent.
\newblock \emph{Privacy, big data, and the public good: Frameworks for engagement}, 1: 44--75.

\bibitem[{Benoit-Bryan, Jean-Mary, and Locks(2023)}]{jen_benoit-bryan_workplace_2023}
Benoit-Bryan, J.; Jean-Mary, D.; and Locks, M. 2023.
\newblock Workplace {Equity} and {Organizational} {Culture} in {US} {Art} {Museums}.
\newblock Technical report, Museums Moving Forward.

\bibitem[{Bergen and Surane(2018)}]{mark_bergen_google_2018}
Bergen, M.; and Surane, J. 2018.
\newblock Google and {Mastercard} {Cut} a {Secret} {Ad} {Deal} to {Track} {Retail} {Sales}.
\newblock \emph{Bloomberg}.

\bibitem[{Boenisch et~al.(2023)Boenisch, Dziedzic, Schuster, Shamsabadi, Shumailov, and Papernot}]{boenisch2023curious}
Boenisch, F.; Dziedzic, A.; Schuster, R.; Shamsabadi, A.~S.; Shumailov, I.; and Papernot, N. 2023.
\newblock When the curious abandon honesty: Federated learning is not private.
\newblock In \emph{2023 IEEE 8th European Symposium on Security and Privacy (EuroS\&P)}, 175--199. IEEE.

\bibitem[{Brown et~al.(2022)Brown, Lee, Mireshghallah, Shokri, and Tram{\`e}r}]{brown2022does}
Brown, H.; Lee, K.; Mireshghallah, F.; Shokri, R.; and Tram{\`e}r, F. 2022.
\newblock What does it mean for a language model to preserve privacy?
\newblock In \emph{Proceedings of the 2022 ACM Conference on Fairness, Accountability, and Transparency}, 2280--2292.

\bibitem[{Burk(2016)}]{burk2016perverse}
Burk, D.~L. 2016.
\newblock Perverse innovation.
\newblock \emph{Wm. \& Mary L. Rev.}, 58: 1.

\bibitem[{Burt, Stalla-Bourdillon, and Rossi(2021)}]{burt2021guide}
Burt, A.; Stalla-Bourdillon, S.; and Rossi, A. 2021.
\newblock A Guide to the EU's Unclear Anonymization Standards.
\newblock \url{https://iapp.org/news/a/a-guide-to-the-eus-unclear-anonymization-standards}.

\bibitem[{{California State Legislature}(2018)}]{ccpa}
{California State Legislature}. 2018.
\newblock 1.81.5. California Consumer Privacy Act of 2018 [1798.100 - 1798.199.100].

\bibitem[{Chang and Shokri(2021)}]{chang2021privacy}
Chang, H.; and Shokri, R. 2021.
\newblock On the privacy risks of algorithmic fairness.
\newblock In \emph{2021 IEEE European Symposium on Security and Privacy (EuroS\&P)}, 292--303. IEEE.

\bibitem[{Chen et~al.(2023)Chen, Nyhan, Reifler, Robertson, and Wilson}]{chen2023subscriptions}
Chen, A.~Y.; Nyhan, B.; Reifler, J.; Robertson, R.~E.; and Wilson, C. 2023.
\newblock Subscriptions and external links help drive resentful users to alternative and extremist YouTube channels.
\newblock \emph{Science Advances}, 9(35): eadd8080.

\bibitem[{Choi(2005)}]{choi2005grokster}
Choi, B.~H. 2005.
\newblock The Grokster Dead-End.
\newblock \emph{Harv. JL \& Tech.}, 19: 393.

\bibitem[{Citron(2022)}]{citron2022fight}
Citron, D. 2022.
\newblock \emph{{The Fight for Privacy: Protecting Dignity, Identity, and Love in the Digital Age}}.
\newblock Random House.

\bibitem[{Corman et~al.(2022)Corman, Canaway, Culnane, and Teague}]{corman_public_2022}
Corman, A.; Canaway, R.; Culnane, C.; and Teague, V. 2022.
\newblock Public comprehension of privacy protections applied to health data shared for research: {An} {Australian} cross-sectional study.
\newblock \emph{International Journal of Medical Informatics}, 167: 104859.

\bibitem[{Cummings and Desai(2018)}]{cummings2018role}
Cummings, R.; and Desai, D. 2018.
\newblock The role of differential privacy in gdpr compliance.
\newblock In \emph{FAT’18: Proceedings of the Conference on Fairness, Accountability, and Transparency}, 20.

\bibitem[{Cummings, Kaptchuk, and Redmiles(2021)}]{cummings_i_2021}
Cummings, R.; Kaptchuk, G.; and Redmiles, E.~M. 2021.
\newblock "{I} need a better description": {An} {Investigation} {Into} {User} {Expectations} {For} {Differential} {Privacy}.
\newblock In \emph{Proceedings of the 2021 {ACM} {SIGSAC} {Conference} on {Computer} and {Communications} {Security}}, 3037--3052. Virtual Event Republic of Korea: ACM.
\newblock ISBN 978-1-4503-8454-4.

\bibitem[{Dandeker(1990)}]{dandeker1990surveillance}
Dandeker, C. 1990.
\newblock \emph{Surveillance, Power, and Modernity: Bureaucracy and Discipline from 1700 to the Present Day}.
\newblock St. Martin's Press.
\newblock ISBN 9780312042226.

\bibitem[{De~Viti et~al.(2022)De~Viti, Sheff, Glaeser, Dinis, Rodrigues, Katz, Bhattacharjee, Hithnawi, Garg, and Druschel}]{de_viti_covault_2022}
De~Viti, R.; Sheff, I.; Glaeser, N.; Dinis, B.; Rodrigues, R.; Katz, J.; Bhattacharjee, B.; Hithnawi, A.; Garg, D.; and Druschel, P. 2022.
\newblock {CoVault}: {A} {Secure} {Analytics} {Platform}.
\newblock \emph{arXiv preprint arXiv:2208.03784}.

\bibitem[{Debenedetti et~al.(2023)Debenedetti, Severi, Carlini, Choquette-Choo, Jagielski, Nasr, Wallace, and Tram{\`e}r}]{debenedetti2023privacy}
Debenedetti, E.; Severi, G.; Carlini, N.; Choquette-Choo, C.~A.; Jagielski, M.; Nasr, M.; Wallace, E.; and Tram{\`e}r, F. 2023.
\newblock Privacy side channels in machine learning systems.
\newblock \emph{arXiv preprint arXiv:2309.05610}.

\bibitem[{{Delaware State Legislature}(2024)}]{delawarepdpa}
{Delaware State Legislature}. 2024.
\newblock Delaware Personal Data Privacy Act.
\newblock \url{https://legiscan.com/DE/text/HB154/id/2807502/Delaware-2023-HB154-Draft.html}.

\bibitem[{Dotan and Milli(2020)}]{dotan2020value}
Dotan, R.; and Milli, S. 2020.
\newblock Value-laden disciplinary shifts in machine learning.
\newblock In \emph{Proceedings of the 2020 Conference on Fairness, Accountability, and Transparency}, 294--294.

\bibitem[{Dwork(2011)}]{dwork2011firm}
Dwork, C. 2011.
\newblock A firm foundation for private data analysis.
\newblock \emph{Communications of the ACM}, 54(1): 86--95.

\bibitem[{Dwork et~al.(2015)Dwork, Feldman, Hardt, Pitassi, Reingold, and Roth}]{dwork2015generalization}
Dwork, C.; Feldman, V.; Hardt, M.; Pitassi, T.; Reingold, O.; and Roth, A. 2015.
\newblock Generalization in adaptive data analysis and holdout reuse.
\newblock \emph{Advances in Neural Information Processing Systems}, 28.

\bibitem[{{EU Legislature}(2018)}]{eugdpr}
{EU Legislature}. 2018.
\newblock General Data Protection Regulation.

\bibitem[{{European Union Legislature}(2022)}]{}
{European Union Legislature}. 2022.
\newblock Digital Services Act.
\newblock \url{https://commission.europa.eu/strategy-and-policy/priorities-2019-2024/europe-fit-digital-age/digital-services-act-ensuring-safe-and-accountable-online-environment_en}.

\bibitem[{Evans, Kolesnikov, and Rosulek(2018)}]{david_evans_pragmatic_2018}
Evans, D.; Kolesnikov, V.; and Rosulek, M. 2018.
\newblock \emph{A {Pragmatic} {Introduction} to {Secure} {Multi}-{Party} {Computation}}.
\newblock NOW Publishers.

\bibitem[{Finck and Pallas(2020)}]{finck2020they}
Finck, M.; and Pallas, F. 2020.
\newblock They who must not be identified—distinguishing personal from non-personal data under the GDPR.
\newblock \emph{International Data Privacy Law}, 10(1): 11--36.

\bibitem[{Forg{\'o}, H{\"a}nold, and Sch{\"u}tze(2017)}]{forgo2017principle}
Forg{\'o}, N.; H{\"a}nold, S.; and Sch{\"u}tze, B. 2017.
\newblock The principle of purpose limitation and big data.
\newblock \emph{New technology, big data and the law}, 17--42.

\bibitem[{Foucault(2012)}]{foucault2012discipline}
Foucault, M. 2012.
\newblock \emph{Discipline and Punish: The Birth of the Prison}.
\newblock Vintage. Knopf Doubleday Publishing Group.
\newblock ISBN 9780307819291.

\bibitem[{Fuchs(2010)}]{fuchs2010can}
Fuchs, C. 2010.
\newblock How Can Surveillance Be Defined? Remarks on Theoretical Foundations.
\newblock \emph{The Internet \& Surveillance - Research Paper Series}.

\bibitem[{Geambasu et~al.(2009)Geambasu, Kohno, Levy, and Levy}]{roxana_geambasu_vanish_2009}
Geambasu, R.; Kohno, T.; Levy, A.~A.; and Levy, H.~M. 2009.
\newblock Vanish: increasing data privacy with self-destructing data.
\newblock In \emph{Proceedings of the 18th {USENIX} {Security} {Symposium}}. Montreal, Canada: USENIX Association.

\bibitem[{Getchell and Varia(2019)}]{ben_getchell_ccd-mpc_2019}
Getchell, B.; and Varia, M. 2019.
\newblock {CCD}-{MPC}.

\bibitem[{Giblin and Ginsburg(2014)}]{giblin2014aereo}
Giblin, R.; and Ginsburg, J.~C. 2014.
\newblock On Aereo and'Avoision'.
\newblock \emph{Copyright Reporter}, 32.

\bibitem[{Gold(2019)}]{gold2019fiduciary}
Gold, A.~S. 2019.
\newblock The Fiduciary Duty of Loyalty.
\newblock \emph{The Oxford Handbook of Fiduciary Law (New York: Oxford University Press, 2019)}.

\bibitem[{Gordon-Tapiero, Wood, and Ligett(2022)}]{gordon2022case}
Gordon-Tapiero, A.; Wood, A.; and Ligett, K. 2022.
\newblock The Case for Establishing a Collective Perspective to Address the Harms of Platform Personalization.
\newblock In \emph{Proceedings of the 2022 Symposium on Computer Science and Law}, 119--130.

\bibitem[{Greenleaf(2020)}]{greenleaf2020california}
Greenleaf, G. 2020.
\newblock California’s CCPA 2.0: Does the US Finally Have a Data Privacy Act?
\newblock \emph{Privacy Laws \& Business International Report}.

\bibitem[{Henderson et~al.(2023)Henderson, Mitchell, Manning, Jurafsky, and Finn}]{henderson2023self}
Henderson, P.; Mitchell, E.; Manning, C.; Jurafsky, D.; and Finn, C. 2023.
\newblock Self-destructing models: Increasing the costs of harmful dual uses of foundation models.
\newblock In \emph{Proceedings of the 2023 AAAI/ACM Conference on AI, Ethics, and Society}, 287--296.

\bibitem[{Hendrix et~al.(2023)Hendrix, Quintin, Sinders, Wylie~Wagner, Bernard, and Mehta}]{justin_hendrix_what_2023}
Hendrix, J.; Quintin, C.; Sinders, C.; Wylie~Wagner, L.; Bernard, T.; and Mehta, A. 2023.
\newblock What is {Secure}? {Analysis} of {Popular} {Messaging} {Apps}.
\newblock Technical report, Tech Policy Press.

\bibitem[{{Illinois State Legislature}(2008)}]{illinoisbipa}
{Illinois State Legislature}. 2008.
\newblock (740 ILCS 14/) Biometric Information Privacy Act.

\bibitem[{Imana, Korolova, and Heidemann(2023)}]{imana2023having}
Imana, B.; Korolova, A.; and Heidemann, J. 2023.
\newblock Having your Privacy Cake and Eating it Too: Platform-supported Auditing of Social Media Algorithms for Public Interest.
\newblock \emph{Proceedings of the ACM on Human-Computer Interaction}, 7(CSCW1): 1--33.

\bibitem[{Jiang et~al.(2021)Jiang, Tan, Peng, Chen, Wu, Zhao, Song, Tong, Liu, Xu et~al.}]{jiang2021gdpr}
Jiang, D.; Tan, C.; Peng, J.; Chen, C.; Wu, X.; Zhao, W.; Song, Y.; Tong, Y.; Liu, C.; Xu, Q.; et~al. 2021.
\newblock A gdpr-compliant ecosystem for speech recognition with transfer, federated, and evolutionary learning.
\newblock \emph{ACM Transactions on Intelligent Systems and Technology (TIST)}, 12(3): 1--19.

\bibitem[{Jones and Tonetti(2020)}]{jones2020nonrivalry}
Jones, C.~I.; and Tonetti, C. 2020.
\newblock Nonrivalry and the Economics of Data.
\newblock \emph{American Economic Review}, 110(9): 2819--2858.

\bibitem[{Kacsmar et~al.(2023)Kacsmar, Duddu, Tilbury, Ur, and Kerschbaum}]{kacsmar_comprehension_2023}
Kacsmar, B.; Duddu, V.; Tilbury, K.; Ur, B.; and Kerschbaum, F. 2023.
\newblock Comprehension from {Chaos}: {Towards} {Informed} {Consent} for {Private} {Computation}.
\newblock In \emph{Proceedings of the 2023 {ACM} {SIGSAC} {Conference} on {Computer} and {Communications} {Security}}, 210--224. Copenhagen Denmark: ACM.
\newblock ISBN 9798400700507.

\bibitem[{Kalluri et~al.(2023)Kalluri, Agnew, Cheng, Owens, Soldaini, and Birhane}]{kalluri2023surveillance}
Kalluri, P.~R.; Agnew, W.; Cheng, M.; Owens, K.; Soldaini, L.; and Birhane, A. 2023.
\newblock The Surveillance AI Pipeline.
\newblock arXiv:2309.15084.

\bibitem[{Kamara(2020)}]{seny_kamara_crypto_2020}
Kamara, S. 2020.
\newblock Crypto for the {People}.

\bibitem[{Kamm et~al.(2013)Kamm, Bogdanov, Laur, and Vilo}]{kamm2013new}
Kamm, L.; Bogdanov, D.; Laur, S.; and Vilo, J. 2013.
\newblock A new way to protect privacy in large-scale genome-wide association studies.
\newblock \emph{Bioinformatics}, 29(7): 886--893.

\bibitem[{Kaptchuk et~al.(2024)Kaptchuk, Benoit-Bryan, Albab, Locks, and Varia}]{gabriel_kaptchuk_jen_benoit-bryan_kinan_dak_albab_mia_locks_mayank_varia_good_2024}
Kaptchuk, G.; Benoit-Bryan, J.; Albab, K.~D.; Locks, M.; and Varia, M. 2024.
\newblock The {Good}, {The} {Bad}, and {The} {Ugly} --- {Lessons} from an {MPC} for {Social} {Good} {Deployment}.

\bibitem[{Karegar, Alaqra, and Fischer-Hübner(2022)}]{karegar_exploring_2022}
Karegar, F.; Alaqra, A.~S.; and Fischer-Hübner, S. 2022.
\newblock Exploring user-suitable metaphors for differentially private data analyses.
\newblock In \emph{Proceedings of the {Eighteenth} {USENIX} {Conference} on {Usable} {Privacy} and {Security}}, {SOUPS}'22. USA: USENIX Association.
\newblock ISBN 978-1-939133-30-4.
\newblock Event-place: Boston, MA, USA.

\bibitem[{Katz(1996)}]{katz1996ill}
Katz, L. 1996.
\newblock \emph{Ill-gotten gains: Evasion, blackmail, fraud, and kindred puzzles of the law}.
\newblock University of Chicago Press.

\bibitem[{Kearns and Roth(2019)}]{kearns2019ethical}
Kearns, M.; and Roth, A. 2019.
\newblock \emph{The ethical algorithm: The science of socially aware algorithm design}.
\newblock Oxford University Press.

\bibitem[{Khan and Pozen(2019)}]{khan2019skeptical}
Khan, L.~M.; and Pozen, D.~E. 2019.
\newblock A skeptical view of information fiduciaries.
\newblock \emph{Harvard Law Review}, 133(2): 497--541.

\bibitem[{Kissner and Song(2005)}]{kissner2005privacy}
Kissner, L.; and Song, D. 2005.
\newblock Privacy-preserving set operations.
\newblock In \emph{Annual International Cryptology Conference}, 241--257. Springer.

\bibitem[{Kone{\v{c}}n{\`y} et~al.(2016)Kone{\v{c}}n{\`y}, McMahan, Ramage, and Richt{\'a}rik}]{konevcny2016federated}
Kone{\v{c}}n{\`y}, J.; McMahan, H.~B.; Ramage, D.; and Richt{\'a}rik, P. 2016.
\newblock Federated optimization: Distributed machine learning for on-device intelligence.
\newblock \emph{arXiv preprint arXiv:1610.02527}.

\bibitem[{Koops(2008)}]{koops2008some}
Koops, B.-J. 2008.
\newblock Some reflections on profiling, power shifts, and protection paradigms.
\newblock \emph{PROFILING THE EUROPEAN CITIZEN, Hildebrandt \& Gutwirth, eds., Springer}.

\bibitem[{Kumar, Hines, and Dickerson(2022)}]{kumar2022equalizing}
Kumar, I.~E.; Hines, K.~E.; and Dickerson, J.~P. 2022.
\newblock Equalizing credit opportunity in algorithms: Aligning algorithmic fairness research with us fair lending regulation.
\newblock In \emph{Proceedings of the 2022 AAAI/ACM Conference on AI, Ethics, and Society}, 357--368.

\bibitem[{Lapets et~al.(2018)Lapets, Jansen, Albab, Issa, Qin, and Varia}]{lapets_accessible_2018}
Lapets, A.; Jansen, F.; Albab, K.~D.; Issa, R.; Qin, L.; and Varia, M. 2018.
\newblock Accessible {Privacy}-{Preserving} {Web}-{Based} {Data} {Analysis} for {Assessing} and {Addressing} {Economic} {Inequalities}.
\newblock In \emph{Proceedings of the 1st {ACM} {SIGCAS} {Conference} on {Computing} and {Sustainable} {Societies}}, 1--5. Menlo Park and San Jose CA USA: ACM.
\newblock ISBN 978-1-4503-5816-3.

\bibitem[{Lessig(2000)}]{lessig2000code}
Lessig, L. 2000.
\newblock Code is law.
\newblock \emph{Harvard magazine}, 1: 2000.

\bibitem[{Ligett and Nissim(2023)}]{ligett2023we}
Ligett, K.; and Nissim, K. 2023.
\newblock We Need to Focus on How Our Data Is Used, Not Just How It Is Shared.
\newblock \emph{Communications of the ACM}, 66(9): 32--34.

\bibitem[{Liu et~al.(2023)Liu, Kang, Zou, Pu, He, Ye, Ouyang, Zhang, and Yang}]{liu2023vertical}
Liu, Y.; Kang, Y.; Zou, T.; Pu, Y.; He, Y.; Ye, X.; Ouyang, Y.; Zhang, Y.-Q.; and Yang, Q. 2023.
\newblock Vertical Federated Learning: Concepts, Advances and Challenges.
\newblock arXiv:2211.12814.

\bibitem[{Lyon(2003)}]{lyon2003surveillance}
Lyon, D. 2003.
\newblock \emph{Surveillance as social sorting: Privacy, risk, and digital discrimination}.
\newblock Psychology Press.

\bibitem[{Lyon(2007)}]{lyon2007surveillance}
Lyon, D. 2007.
\newblock \emph{Surveillance studies: An overview}.
\newblock Polity.

\bibitem[{Malmgren(2019)}]{malmgren2019resisting}
Malmgren, E. 2019.
\newblock Resisting ``Big Other'': What Will It Take to Defeat Surveillance Capitalism?
\newblock In \emph{New Labor Forum}, volume~28, 42--50. SAGE Publications Sage CA: Los Angeles, CA.

\bibitem[{Mammen(2021)}]{mammen2021federated}
Mammen, P.~M. 2021.
\newblock Federated learning: Opportunities and challenges.
\newblock \emph{arXiv preprint arXiv:2101.05428}.

\bibitem[{McGuigan et~al.(2023)McGuigan, Sivan-Sevilla, Parham, and Shvartzshnaider}]{mcguigan2023private}
McGuigan, L.; Sivan-Sevilla, I.; Parham, P.; and Shvartzshnaider, Y. 2023.
\newblock Private attributes: The meanings and mechanisms of “privacy-preserving” adtech.
\newblock \emph{new media \& society}, 14614448231213267.

\bibitem[{Microsoft(2023)}]{microsoftcompliance2023}
Microsoft. 2023.
\newblock Federated Learning with Azure Machine Learning: Powering Privacy-Preserving Innovation in AI.
\newblock \url{https://techcommunity.microsoft.com/t5/ai-machine-learning-blog/federated-learning-with-azure-machine-learning-powering-privacy/ba-p/3824720}.

\bibitem[{Miller(2019)}]{miller2018identification}
Miller, P.~B. 2019.
\newblock The identification of fiduciary relationships.
\newblock \emph{The Oxford Handbook of Fiduciary Law (New York: Oxford University Press, 2019)}.

\bibitem[{Nanayakkara et~al.(2023)Nanayakkara, Smart, Cummings, Kaptchuk, and Redmiles}]{nanayakkara_what_2023}
Nanayakkara, P.; Smart, M.~A.; Cummings, R.; Kaptchuk, G.; and Redmiles, E.~M. 2023.
\newblock What {Are} the {Chances}? {Explaining} the {Epsilon} {Parameter} in {Differential} {Privacy}.
\newblock In \emph{32nd {USENIX} {Security} {Symposium} ({USENIX} {Security} 23)}, 1613--1630. Anaheim, CA: USENIX Association.
\newblock ISBN 978-1-939133-37-3.

\bibitem[{{New Jersey State Legislature}(2023)}]{newjerseysb332}
{New Jersey State Legislature}. 2023.
\newblock SB332.
\newblock \url{https://www.njleg.state.nj.us/bill-search/2022/S332}.

\bibitem[{Nissenbaum(1996)}]{nissenbaum1996accountability}
Nissenbaum, H. 1996.
\newblock Accountability in a computerized society.
\newblock \emph{Science and engineering ethics}, 2: 25--42.

\bibitem[{Nissenbaum(2004)}]{nissenbaum2004privacy}
Nissenbaum, H. 2004.
\newblock Privacy as contextual integrity.
\newblock \emph{Wash. L. Rev.}, 79: 119.

\bibitem[{Noyb(2024)}]{noybmicrosoft2024}
Noyb. 2024.
\newblock Microsoft violates children’s privacy – but blames your local school.
\newblock \url{https://noyb.eu/en/microsoft-violates-childrens-privacy-blames-your-local-school}.

\bibitem[{Papernot et~al.(2018)Papernot, McDaniel, Sinha, and Wellman}]{papernot2018sok}
Papernot, N.; McDaniel, P.; Sinha, A.; and Wellman, M.~P. 2018.
\newblock Sok: Security and privacy in machine learning.
\newblock In \emph{2018 IEEE European Symposium on Security and Privacy (EuroS\&P)}, 399--414. IEEE.

\bibitem[{Papernot and Thakurta(2021)}]{nisthowto2021}
Papernot, N.; and Thakurta, G.~A. 2021.
\newblock How to deploy machine learning with differential privacy.

\bibitem[{Parsons and Viljoen(2023)}]{parsons2023valuing}
Parsons, A.; and Viljoen, S. 2023.
\newblock Valuing Social Data.
\newblock \emph{Columbia Law Review}, 124.

\bibitem[{Party(2014)}]{wp}
Party, A. . D. P.~W. 2014.
\newblock Opinion 05/2014 on Anonymisation Techniques.
\newblock \url{https://ec.europa.eu/justice/article-29/documentation/opinion-recommendation/files/2014/wp216_en.pdf}.

\bibitem[{Perlman(2005)}]{perlman2005ephemerizer}
Perlman, R. 2005.
\newblock The ephemerizer: Making data disappear.

\bibitem[{Pollman(2019)}]{pollman2019tech}
Pollman, E. 2019.
\newblock Tech, regulatory arbitrage, and limits.
\newblock \emph{European Business organization law review}, 20: 567--590.

\bibitem[{Ponomareva et~al.(2023)Ponomareva, Hazimeh, Kurakin, Xu, Denison, McMahan, Vassilvitskii, Chien, and Thakurta}]{ponomareva2023dp}
Ponomareva, N.; Hazimeh, H.; Kurakin, A.; Xu, Z.; Denison, C.; McMahan, H.~B.; Vassilvitskii, S.; Chien, S.; and Thakurta, A.~G. 2023.
\newblock How to dp-fy ml: A practical guide to machine learning with differential privacy.
\newblock \emph{Journal of Artificial Intelligence Research}, 77: 1113--1201.

\bibitem[{Pozen(2016)}]{pozen2016privacy}
Pozen, D.~E. 2016.
\newblock Privacy-privacy tradeoffs.
\newblock \emph{The University of Chicago Law Review}, 221--247.

\bibitem[{Qin et~al.(2019)Qin, Lapets, Jansen, Flockhart, Albab, Globus-Harris, Roberts, and Varia}]{lucy_qin_usability_2019}
Qin, L.; Lapets, A.; Jansen, F.; Flockhart, P.; Albab, K.~D.; Globus-Harris, I.; Roberts, S.; and Varia, M. 2019.
\newblock From Usability to Secure Computing and Back Again.
\newblock In \emph{Fifteenth Symposium on Usable Privacy and Security (SOUPS 2019)}, 191--210. Santa Clara, CA: USENIX Association.
\newblock ISBN 978-1-939133-05-2.

\bibitem[{Reyes(2022)}]{james_reyes_building_2022}
Reyes, J. 2022.
\newblock Building the next generation of digital advertising in {MPC}.

\bibitem[{Richards and Hartzog(2021)}]{richards2021duty}
Richards, N.; and Hartzog, W. 2021.
\newblock A duty of loyalty for privacy law.
\newblock \emph{Wash. UL Rev.}, 99: 961.

\bibitem[{Sagar and Keke(2021)}]{sagar2021confidential}
Sagar, S.; and Keke, C. 2021.
\newblock Confidential machine learning on untrusted platforms: a survey.
\newblock \emph{Cybersecurity}, 4(1): 1--19.

\bibitem[{Schneier(2023)}]{schneier2023hacker}
Schneier, B. 2023.
\newblock \emph{A Hacker's Mind: How the Powerful Bend Society's Rules, and how to Bend Them Back}.
\newblock WW Norton \& Company.

\bibitem[{Schwartz and Solove(2011)}]{schwartz2011pii}
Schwartz, P.~M.; and Solove, D.~J. 2011.
\newblock The PII problem: Privacy and a new concept of personally identifiable information.
\newblock \emph{NYUL rev.}, 86: 1814.

\bibitem[{Solove(2023)}]{solove2023data}
Solove, D.~J. 2023.
\newblock Data is what data does: Regulating Use, Harm, and risk instead of sensitive data.
\newblock \emph{118 Northwestern University Law Review (Forthcoming)}.

\bibitem[{Solow-Niederman(2022)}]{solow2022information}
Solow-Niederman, A. 2022.
\newblock Information privacy and the inference economy.
\newblock \emph{Nw. UL Rev.}, 117: 357.

\bibitem[{Staab et~al.(2023)Staab, Vero, Balunovi{\'c}, and Vechev}]{staab2023beyond}
Staab, R.; Vero, M.; Balunovi{\'c}, M.; and Vechev, M. 2023.
\newblock Beyond memorization: Violating privacy via inference with large language models.
\newblock \emph{arXiv preprint arXiv:2310.07298}.

\bibitem[{Stojkovic et~al.(2022)Stojkovic, Woodbridge, Fang, Cai, Petrov, Iyer, Huang, Yau, Kumar, Jawa et~al.}]{stojkovic2022applied}
Stojkovic, B.; Woodbridge, J.; Fang, Z.; Cai, J.; Petrov, A.; Iyer, S.; Huang, D.; Yau, P.; Kumar, A.~S.; Jawa, H.; et~al. 2022.
\newblock Applied federated learning: Architectural design for robust and efficient learning in privacy aware settings.
\newblock \emph{arXiv preprint arXiv:2206.00807}.

\bibitem[{Su and Khoshgoftaar(2009)}]{su2009survey}
Su, X.; and Khoshgoftaar, T.~M. 2009.
\newblock A survey of collaborative filtering techniques.
\newblock \emph{Advances in artificial intelligence}, 2009.

\bibitem[{Suriyakumar et~al.(2021)Suriyakumar, Papernot, Goldenberg, and Ghassemi}]{suriyakumar2021chasing}
Suriyakumar, V.~M.; Papernot, N.; Goldenberg, A.; and Ghassemi, M. 2021.
\newblock Chasing your long tails: Differentially private prediction in health care settings.
\newblock In \emph{Proceedings of the 2021 ACM Conference on Fairness, Accountability, and Transparency}, 723--734.

\bibitem[{{Texas State Legislature}(2023)}]{texasdataprivacyandsecurityact}
{Texas State Legislature}. 2023.
\newblock Texas Data Privacy and Security Act.
\newblock \url{https://capitol.texas.gov/BillLookup/Text.aspx?LegSess=88R&Bill=HB4}.

\bibitem[{Treiber et~al.(2022)Treiber, M{\"u}llmann, Schneider, and Spiecker~genannt D{\"o}hmann}]{treiber2022data}
Treiber, A.; M{\"u}llmann, D.; Schneider, T.; and Spiecker~genannt D{\"o}hmann, I. 2022.
\newblock Data Protection Law and Multi-Party Computation: Applications to Information Exchange between Law Enforcement Agencies.
\newblock In \emph{Proceedings of the 21st Workshop on Privacy in the Electronic Society}, 69--82.

\bibitem[{{United States Congress}(2021)}]{datacareact}
{United States Congress}. 2021.
\newblock S.919 - Data Care Act of 2021.
\newblock \url{https://www.congress.gov/bill/117th-congress/senate-bill/919}.

\bibitem[{{U.S. Legislature}(2022)}]{adppa}
{U.S. Legislature}. 2022.
\newblock H.R.8152 - American Data Privacy and Protection Act.

\bibitem[{Van~Loo(2019)}]{van2019regulatory}
Van~Loo, R. 2019.
\newblock Regulatory Monitors.
\newblock \emph{Columbia Law Review}, 119(2): 369--444.

\bibitem[{Van~Loo(2022)}]{van2022privacy}
Van~Loo, R. 2022.
\newblock Privacy Pretexts.
\newblock \emph{Cornell L. Rev.}, 108: 1.

\bibitem[{Veale(2023)}]{veale2023rights}
Veale, M. 2023.
\newblock Rights for Those Who Unwillingly, Unknowingly and Unidentifiably Compute!
\newblock \emph{The Person and the Future of Private Law}.

\bibitem[{Verge(2023)}]{23andme2023}
Verge, T. 2023.
\newblock 23andMe admits hackers accessed 6.9 million users’ DNA Relatives data.
\newblock \url{https://www.theverge.com/2023/12/4/23988050/23andme-hackers-accessed-user-data-confirmed}.

\bibitem[{Viljoen(2021)}]{viljoen2021relational}
Viljoen, S. 2021.
\newblock A relational theory of data governance.
\newblock \emph{Yale LJ}, 131: 573.

\bibitem[{Visa(2022)}]{visasecure2022}
Visa. 2022.
\newblock Secure Collaborative Machine Learning.
\newblock \url{https://usa.visa.com/dam/VCOM/regional/na/us/about-visa/research/documents/secure-collaborative-machine-learning.pdf}.

\bibitem[{Volgushev et~al.(2019)Volgushev, Schwarzkopf, Getchell, Varia, Lapets, and Bestavros}]{volgushev_conclave_2019}
Volgushev, N.; Schwarzkopf, M.; Getchell, B.; Varia, M.; Lapets, A.; and Bestavros, A. 2019.
\newblock Conclave: secure multi-party computation on big data.
\newblock In \emph{Proceedings of the {Fourteenth} {EuroSys} {Conference} 2019}, 1--18. Dresden Germany: ACM.
\newblock ISBN 978-1-4503-6281-8.

\bibitem[{Walsh et~al.(2022)Walsh, Varia, Cohen, Sellars, and Bestavros}]{walsh2022multi}
Walsh, J.~M.; Varia, M.; Cohen, A.; Sellars, A.; and Bestavros, A. 2022.
\newblock Multi-Regulation Computing: Examining the Legal and Policy Questions That Arise From Secure Multiparty Computation.
\newblock In \emph{Proceedings of the 2022 Symposium on Computer Science and Law}, 53--65.

\bibitem[{Wang, Ko, and Mickens(2019)}]{wang2019riverbed}
Wang, F.; Ko, R.; and Mickens, J. 2019.
\newblock Riverbed: Enforcing user-defined privacy constraints in distributed web services.
\newblock In \emph{16th USENIX Symposium on Networked Systems Design and Implementation (NSDI 19)}, 615--630.

\bibitem[{Wang et~al.(2024)Wang, Viti, Dubey, and Redmiles}]{wang_role_2024}
Wang, R.; Viti, R.~D.; Dubey, A.; and Redmiles, E.~M. 2024.
\newblock The {Role} of {Privacy} {Guarantees} in {Voluntary} {Donation} of {Private} {Data} for {Altruistic} {Goals}.
\newblock \_eprint: 2407.03451.

\bibitem[{Wu(2003)}]{wu2003code}
Wu, T. 2003.
\newblock When code isn't law.
\newblock \emph{Virginia Law Review}, 679--751.

\bibitem[{Xiang(2022)}]{xiang2022being}
Xiang, A. 2022.
\newblock Being'Seen'vs.'Mis-Seen': Tensions between Privacy and Fairness in Computer Vision.
\newblock \emph{Harvard Journal of Law \& Technology}.

\bibitem[{Xin et~al.(2022)Xin, Geng, Hu, Chen, Yang, Wang, and Huang}]{xin2022federated}
Xin, B.; Geng, Y.; Hu, T.; Chen, S.; Yang, W.; Wang, S.; and Huang, L. 2022.
\newblock Federated synthetic data generation with differential privacy.
\newblock \emph{Neurocomputing}, 468: 1--10.

\bibitem[{Xiong et~al.(2023)Xiong, Wu, Wang, Proctor, Blocki, Li, and Jha}]{xiong_exploring_2023}
Xiong, A.; Wu, C.; Wang, T.; Proctor, R.~W.; Blocki, J.; Li, N.; and Jha, S. 2023.
\newblock Exploring {Use} of {Explanative} {Illustrations} to {Communicate} {Differential} {Privacy} {Models}.
\newblock \emph{Proceedings of the Human Factors and Ergonomics Society Annual Meeting}, 67(1): 226--232.

\bibitem[{Xu, Baracaldo, and Joshi(2021)}]{xu2021privacy}
Xu, R.; Baracaldo, N.; and Joshi, J. 2021.
\newblock Privacy-preserving machine learning: Methods, challenges and directions.
\newblock \emph{arXiv preprint arXiv:2108.04417}.

\bibitem[{Yew and Xiang(2022)}]{yew2022regulating}
Yew, R.-J.; and Xiang, A. 2022.
\newblock Regulating Facial Processing Technologies: Tensions Between Legal and Technical Considerations in the Application of Illinois BIPA.
\newblock In \emph{2022 ACM Conference on Fairness, Accountability, and Transparency}, 1017--1027.

\bibitem[{Zarsky(2019)}]{zarsky2019privacy}
Zarsky, T.~Z. 2019.
\newblock Privacy and manipulation in the digital age.
\newblock \emph{Theoretical Inquiries in Law}, 20(1): 157--188.

\bibitem[{Zeng et~al.(2010)Zeng, Shi, Xu, and Feng}]{zeng2010safevanish}
Zeng, L.; Shi, Z.; Xu, S.; and Feng, D. 2010.
\newblock Safevanish: An improved data self-destruction for protecting data privacy.
\newblock In \emph{2010 IEEE Second International Conference on Cloud Computing Technology and Science}, 521--528. IEEE.

\bibitem[{Zhang et~al.(2021)Zhang, Xie, Bai, Yu, Li, and Gao}]{zhang2021survey}
Zhang, C.; Xie, Y.; Bai, H.; Yu, B.; Li, W.; and Gao, Y. 2021.
\newblock A survey on federated learning.
\newblock \emph{Knowledge-Based Systems}, 216: 106775.

\bibitem[{Zhong and Bu(2022)}]{zhong2022privacy}
Zhong, H.; and Bu, K. 2022.
\newblock Privacy-utility trade-off.
\newblock \emph{arXiv preprint arXiv:2204.12057}.

\end{thebibliography}

%%%%%%%%%%%%%%%%%%%%%%%%%%%%%%%%%%%%%%%%%%%%%%%%%%%%%%%%%%%%%%%%%%%%%%%%%%%%%%%%
\end{document}